\documentclass[]{aastex631}
\usepackage{amsmath}
\usepackage{booktabs}
\usepackage{comment}
\usepackage{xspace}
\usepackage{enumitem}
\usepackage{chronosys}
\usepackage{longtable}
\usepackage{latexsym}
\usepackage{amssymb}
\usepackage{txfonts}
\usepackage{longtable}
\usepackage{graphicx}
\usepackage{epsf}
\usepackage{url}
\usepackage{hyperref}

\newcommand\src{EP240709A\xspace}

\begin{document}

\title{Tentative Blazar Candidate EP240709A Associated with 4FGL~J0031.5$-$5648: NICER and Archival Multiwavelength Observations}

\correspondingauthor{Mason Ng}
\email{mason.ng@mcgill.ca}

\author[0000-0002-0940-6563]{Mason Ng}
\affiliation{MIT Kavli Institute for Astrophysics and Space Research, Massachusetts Institute of Technology, Cambridge, MA 02139, USA}
\affiliation{Department of Physics, McGill University, 3600 rue University, Montr\'{e}al, QC H3A 2T8, Canada}
\affiliation{Trottier Space Institute, McGill University, 3550 rue University, Montr\'{e}al, QC H3A 2A7, Canada}

\author[0000-0002-8548-482X]{Jeremy Hare}
\affiliation{Astrophysics Science Division, NASA Goddard Space Flight Center, Greenbelt, MD 20771, USA}
\affiliation{Center for Research and Exploration in Space Science \& Technology II (CRESST II), NASA/GSFC, Greenbelt, MD 20771, USA}
\affiliation{The Catholic University of America, 620 Michigan Ave., N.E. Washington, DC 20064, USA}

\author[0000-0002-6789-2723]{Gaurava K. Jaisawal}
\affiliation{DTU Space, Technical University of Denmark, Elektrovej 327-328, DK-2800 Lyngby, Denmark}

\author[0000-0002-0380-0041]{Christian Malacaria}
\affiliation{INAF Osservatorio Astronomico di Roma, Via Frascati 33, 00078 Monte Porzio Catone (RM), Italy}

\author[0000-0001-9803-3879]{Craig B. Markwardt}
\affiliation{Astrophysics Science Division, NASA Goddard Space Flight Center, Greenbelt, MD 20771, USA}

\author[0000-0002-0118-2649]{Andrea Sanna}
\affiliation{Dipartimento di Fisica, Universit\`a degli Studi di Cagliari, SP Monserrato-Sestu km 0.7, I-09042 Monserrato, Italy}




\begin{abstract}

We report on follow-up observations of the recently discovered transient by the Einstein Probe, \src, with the Neutron star Interior Composition Explorer (NICER). We also incorporated archival multiwavelength survey data from the Neil Gehrels Swift Observatory (X-ray), Gaia (optical), the Fermi Gamma-ray Space Telescope (gamma-ray), and the Wide-field Infrared Survey Explorer (infrared) to distinguish between blazars and stellar systems. We suggest that \src is likely an active blazar.

\end{abstract}

\keywords{Blazars (164) --- Active galactic nuclei (16)}


\section{Introduction} \label{sec:intro}

\src was first discovered by the Wide-field X-ray Telescope (WXT) aboard the Einstein Probe \citep[EP;][]{yuan22} on 2024 July 9 with a 0.5--4.0~keV unabsorbed flux of $1.3_{-0.5}^{+0.6}\times10^{-11}{\rm\,erg\,s^{-1}\,cm^{-2}}$ \citep{atel16704}. A follow-up 4~ks observation by the Follow-up X-ray Telescope (FXT) onboard EP on 2024 July 11 showed that the source did not exhibit significant variability \citep{atel16704}. Finally, archival Swift/XRT and eROSITA observations suggested that \src is likely associated with 4FGL~J0031.5$-$5648, due to the overlapping X-ray localization regions (see Section~\ref{sec:mw}) between Swift/XRT, eROSITA, and EP/FXT \citep{atel16704}. In this article, we report on follow-up NICER X-ray observations and utilize multiwavelength survey data to constrain the possible nature of the source.


\section{Observations}

The Neutron Star Interior Composition Explorer (NICER) observed \src from 2024 July 17 to 2024 July 26 (ObsIDs 720467010x) for a total filtered exposure of 10.4~ks. The data were processed with \texttt{nicerl2} with the default filtering criteria from \texttt{2024-02-09\_V012A} of the NICER Data Analysis Software (\texttt{NICERDAS}) available through \texttt{HEASoft} 6.33. For our pulsation searches and aperiodic timing analysis, we corrected the photon arrival times to the Solar System barycenter with the JPL DE421 solar system ephemeris \citep{folkner09} and with source coordinates ${\rm R.A.} = 7\fdg8964$, ${\rm DEC.} = -56\fdg7787$ (position error of $3.6\arcsec$) from the Living Swift XRT Point Source Catalogue \citep[LSXPS;][]{evans23}.

We generated spectral products with the \texttt{nicerl3-spect} pipeline, and rebinned the spectra with optimal binning \citep{kaastra16} and ensured 25 minimum counts per spectral bin. We utilized the \texttt{nibackgen3C50} model \citep{remillard22} to describe the background. Since the background spectra dominated above 3~keV, we restricted the spectral fitting to the 0.4--3.0~keV energy range. We did not use data from ObsID 7204670106 due to the short exposure (22~s). The source and background (\texttt{sw} model\footnote{\url{https://heasarc.gsfc.nasa.gov/docs/nicer/analysis_threads/background/}}) light curves were constructed with \texttt{nicerl3-lc}.  

\section{Results}

\subsection{NICER}

The light curve (see Figure~\ref{fig:source} a) does not show significant flaring activity. Inspecting the light curve with finer time bins (1~s bins; not plotted), we did not see any variability that would suggest a transitional millisecond pulsar in the sub-luminous disk state \citep{papitto22}.

We searched for coherent pulsations with acceleration searches \citep{ransom02} over each of the 24 good time intervals in the 0.4--3.0~keV band. Acceleration searches are most sensitive for time intervals $T < P_{\rm orb}/10$, where $T$ is the interval length and $P_{\rm orb}$ is the binary orbital period \citep{ransom02}. The minimum, maximum, and median interval lengths were 430~s, 22~s, and 1077~s, respectively. We did not see any significant signals. Given the relative faintness of the source (less than 4~c/s), we caution interpreting non-detections in short (less than 500~s) intervals. We also constructed the 0.4--3.0~keV averaged power spectrum with 128~s bins to search for quasi-periodic oscillations, but we found a power spectrum consistent with Poisson noise. We place upper limits on the sinusoidal amplitude of any pulsed signal at $\sim31\%$ at 1~Hz and $\sim35\%$ at 500~Hz using the formalism described in \cite{vaughan94}. 

The NICER spectra were well described by an absorbed power law (reduced $\chi^2 \sim1.0$ for $\sim35$~d.o.f.). The power law index, $\Gamma$, varied between 1.9 and 2.6, and the 0.4--3.0~keV unabsorbed flux varied between (3--9)$\times10^{-12}{\rm\,erg\,s^{-1}\,cm^{-2}}$, indicating a slight decrease from the flux level initially reported by the EP team \citep{atel16704}. The spectral parameters are also shown in Figures \ref{fig:source} a), b), and c). We fixed $n_H = 0.048\times10^{22}{\rm\,cm^{-2}}$, an exposure time-weighted average value taken from initially fitting an absorbed power law to all spectra from the NICER observations (with $n_H$ being free).

\subsection{Multiwavelength Counterparts} \label{sec:mw}

The initial discovery of \src by the EP team associated it with 4FGL~J0031.5$-$5648, an unassociated Fermi source. In Figures~\ref{fig:source} d) and e), we plot the variability indices, curvature significances, as well as the curvature parameters and peak energies (MeV) for three source classes: blazars, pulsars, and unassociated sources. The variability index quantifies the temporal variability of the source, and the curvature significance quantifies the spectral fit improvement between a power law and a log parabola (LP). The LP model \citep{zheng23} is described by,
\begin{equation}
    \frac{dN}{dE} = N_0\left(\frac{E}{E_b}^{-\left[\alpha+\beta\text{log}\left(\frac{E}{E_b}\right)\right]}\right),
\end{equation}
where $N_0$ is the normalization constant, $E_b$ is the pivot energy (MeV), $\alpha$ is the photon index at the pivot energy, and $\beta$ is the curvature parameter. The peak energy\footnote{\url{https://docs.gammapy.org/0.7/api/gammapy.spectrum.models.LogParabola.html}} is given by $E_{\rm peak} = E_b\text{exp}[(2-\alpha)/(2\beta)]$. We made use of data from the first version of the Fermi Large Area Telescope 14-year Source Catalog (\texttt{gll\_psc\_v32}). We also plot the unassociated Fermi source, 4FGL~J0031.5$-$5648, on the figures with a magenta triangle.

There are several possible multiwavelength counterparts that lie within the Swift LSXPS and EP/FXT error regions. The Gaia counterpart, Gaia EDR3 4918773190794845824, has source coordinates ${\rm R.A.} = 7\fdg89783331(7)$, ${\rm DEC.} = -56\fdg77818634(7)$ and a proper motion consistent with zero, where ${\rm PMRA} = 0.2\pm0.3{\rm\,mas/yr}$ and ${\rm PMDEC} = -0.07\pm0.33{\rm\,mas/yr}$. The eROSITA counterpart was pointed out by \cite{atel16704}, with ${\rm R.A.} = 7\fdg8973(4)$, ${\rm DEC.} = -56\fdg7772(5)$ and IAU name 1eRASS~J003135.3$-$564638 \citep{merloni24}. We also identified an infrared counterpart with source coordinates ${\rm R.A.} = 7\fdg89771(6)$, ${\rm DEC.} = -56\fdg77815(6)$, which is listed in the Wide-field Infrared Survey Explorer \citep[WISE;][]{wright10} All-Sky Source Catalog from the WISE satellite \citep{wisecatalog}, which surveyed the mid-infrared sky once in the four bands centered on 3.4, 4.6, 12, and 22~$\mu{\rm m}$. The catalog listed significant measurements for the first three bands, with magnitudes $W_1 = 15.38(4)$, $W_2 = 15.14(9)$, and $W_3=12.6(4)$, respectively. We can thus determine WISE colors that will aid in classifying the source, where $W_{12}\coloneqq W_1-W_2 = 0.24(10)$ and $W_{23}\coloneqq W_2-W_3 = 2.6(4)$.

\section{Discussion}

\begin{figure}[htbp!]
\plotone{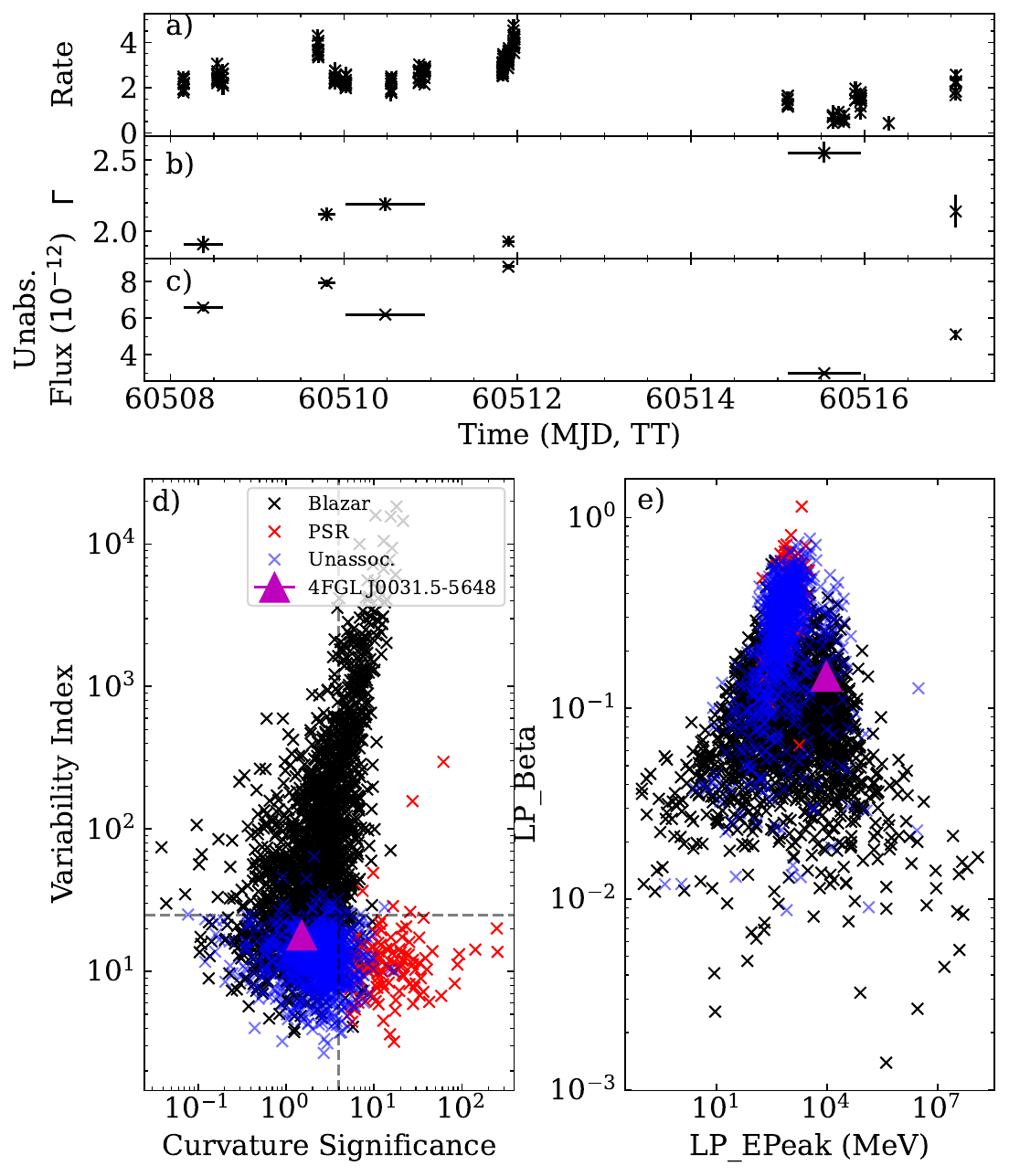}
\caption{a) 0.4--3.0~keV background-subtracted NICER light curve (64~s bins); b) Power law photon index; and c) 0.3--10.0~keV unabsorbed flux in units of $10^{-12}{\rm\,erg\,s^{-1}\,cm^{-2}}$. d) variability index vs. curvature significance; and e) curvature parameter and peak energy (in MeV). 
\label{fig:source}}
\end{figure}

We suggest that \src is likely an active blazar from several independent pieces of information: 
\begin{itemize}
    \item The X-ray spectral and flux variability as shown in Figures~\ref{fig:source} a), b), and c) are consistent with that of AGN \citep{mushotzky93,mchardy10}.
    \item The diagnostics shown in Figures~\ref{fig:source} d) and e) show that 4FGL~J0031.5$-$5648 is more likely to be a blazar compared to a pulsar, as it has a lower curvature significance, lower curvature parameter, and higher peak energy than pulsars in general.
    \item The machine-learning classification for the Gaia Data Release 3 (Gaia DR3) data set labeled a 99.98\% probability for the source to be a quasar, determined through the Gaia Discrete Source Classifier utilizing Gaussian mixture models \citep{delchambre23}. We could not find any spectra from the Gaia archive.
    \item The two WISE colors are consistent with the infrared counterpart being a blazar, based on the $W_1$-$W_2$-$W_3$ $\mu$m color-color diagram of WISE thermal sources and blazars \citep{massaro11,abrusco19}. The counterpart could also be a high-frequency-peaked BL Lac \citep{massaro11}. Stellar systems tend to have redder $W_1$-$W_2$ (below 0.0) and $W_2$-$W_3$ (below 1.0) colors \citep{abrusco19}.
\end{itemize}
 
We note that given the large error circle of 4FGL~J0031.5$-$5648 of roughly 4$\arcmin$ in the Fermi catalog, the counterparts identified above could simply be spatially coincident. 

\vspace{5mm}
\facilities{NICER, eROSITA\footnote{\url{https://erosita.mpe.mpg.de/dr1/AllSkySurveyData_dr1/acknowledgement.html}\label{erosita_acknow}}, Swift, Gaia, Fermi, WISE\footnote{\url{https://wise2.ipac.caltech.edu/docs/release/allsky/expsup/sec1_6b.html}\label{wise_acknow}}}


\software{Astropy \citep{astropy:2013, astropy:2018}, NumPy and SciPy \citep{virtanen20}, Matplotlib \citep{hunter07}, IPython \citep{perez07}, HEASoft 6.33\footnote{\url{https://heasarc.gsfc.nasa.gov/docs/heasarc/acknow.html}\label{heasoft_acknow}} \citep{heasoft}}



\begin{acknowledgments}

Due to the word limit, we truncate the acknowledgements and link to the full text. This research has made use of data and/or software provided by HEASARC\textsuperscript{\ref{heasoft_acknow}}. This work is based on data from eROSITA, the soft X-ray instrument aboard SRG\textsuperscript{\ref{erosita_acknow}}, and from the Wide-field Infrared Survey Explorer\textsuperscript{\ref{wise_acknow}}. M.N. is a Fonds de Recherche du Quebec – Nature et Technologies (FRQNT) postdoctoral fellow.

\end{acknowledgments}



\bibliography{ulx}{}
\bibliographystyle{aasjournal}

\end{document}